\documentclass[12pt]{article}

\def\gtwid{\mathrel{\raise.3ex\hbox{$>$\kern-.75em\lower1ex\hbox{$\sim$}}}}
\def\ltwid{\mathrel{\raise.3ex\hbox{$<$\kern-.75em\lower1ex\hbox{$\sim$}}}}
\def\square{\kern1pt\vbox{\hrule height 1.2pt\hbox{\vrule width 1.2pt\hskip 3pt
   \vbox{\vskip 6pt}\hskip 3pt\vrule width 0.6pt}\hrule height 0.6pt}\kern1pt}

\begin{document}

\begin{titlepage}

\begin{flushright}
BRX-TH-6689 \\
Calt-68-2946 \\
UFIFT-QG-13-05
\end{flushright}

\vskip 2cm

\begin{center}
{\bf Observational Viability and Stability of Nonlocal Cosmology}
\end{center}

\vskip 2cm

\begin{center}
S. Deser$^*$
\end{center}

\begin{center}
\it{California Institute of Technology, Pasadena, CA 91125 and \\
Department of Physics, Brandeis University, Waltham, MA 02254}
\end{center}

\begin{center}
and
\end{center}

\begin{center}
R. P. Woodard$^{\dagger}$
\end{center}

\begin{center}
\it{Department of Physics, University of Florida, Gainesville, FL 32611}
\end{center}

\vspace{1cm}

\begin{center}
ABSTRACT
\end{center}
We show that the nonlocal gravity models, proposed to explain
current cosmic acceleration without dark energy, pass two major
tests: First, they can be defined so as not to alter the,
observationally correct, general relativity predictions for
gravitationally bound systems. Second, they are stable, ghost-free,
with no additional excitations beyond those of general relativity.
In this they differ from their, ghostful, localized versions. The
systems' initial value constraints are the same as in general
relativity, and our nonlocal modifications never convert the
original gravitons into ghosts.

\begin{flushleft}
PACS numbers: 95.36.+x, 04.50.Kd, 11.10.Lm
\end{flushleft}

\begin{flushleft}
$^*$ e-mail: deser@brandeis.edu \\
$^{\dagger}$ e-mail: woodard@phys.ufl.edu
\end{flushleft}

\end{titlepage}

\section{Introduction}

Explaining the current phase of cosmic acceleration is an ongoing
challenge \cite{reviews}. The data are consistent with general
relativity operating on a critical energy density whose current
composition is about 70\% cosmological constant plus about 30\%
nonrelativistic, and small amounts of relativistic, matter
\cite{Yun,Yunbook}. However, there is no good explanation for why
the cosmological constant should be so small, nor why it should
recently have come into dominance \cite{Lambdarevs}. Scalar
potential models \cite{earlyquint,Paul} can be devised to reproduce
the observed expansion history \cite{TW1,scalrecon} but they must be
fine tuned and are difficult to motivate. Quantum effects from a
very light scalar have also been suggested \cite{Parker}.

Various modifications of general relativity that generalize its
Lagrangian from $R$ to $f(R)$ \cite{Mark,NO} represent the only
local, metric-based, generally coordinate invariant and stable
modification of gravity \cite{RPW}. But the sole model within this
class that exactly reproduces the $\Lambda$CDM expansion history is
general relativity with $f(R) = R - 2 \Lambda$ \cite{nof(R)}.

More modification freedom is available if locality is abandoned
\cite{nonloc}, but this novel territory raises the worry of new
degrees of freedom (DoF), possibly of instability-negative energy
\cite{EW}. While we do not believe such models to be fundamental,
even if observationally viable in some regime of validity, they must
still face the above problems of principle, as well as more
phenomenological ones. Their origin would be the gravitational
corrections that grew non-perturbatively during the primordial
inflation epoch \cite{TW1}, a conjecture that, while plausible
\cite{SY,TW2}, is as yet unverified \cite{MW}. Independent of their
ultimate origin, these models have been proposed and studied purely
phenomenologically. Ours \cite{ourmodel} adds the nonlocal piece
\begin{equation}
\Delta \mathcal{L} \equiv \frac1{16 \pi G} \, R \sqrt{-g} \times
f\Bigl( \frac1{\square} \, R \Bigr) \; , \label{DL2}
\end{equation}
to the Einstein term $R \sqrt{-g}/16\pi G$. Our signature is
$(-+++)$, with the convention $R_{\mu\nu} \sim +\partial_{\rho}
\Gamma^{\rho}_{~ \mu\nu}$. The inverse of the (scalar) d`Alembertian
$\square \equiv (-g)^{-\frac12} \partial_{\mu} [\sqrt{-g} \,
g^{\mu\nu} \partial_{\nu}]$ is the retarded one, with vanishing 0th
and 1st time derivatives at the initial time \cite{ourmodel}. In
addition to simplicity, the great advantage of this class of models
is to provide a natural delay for the onset of cosmic acceleration:
because the Ricci scalar $R$ vanishes during radiation dominance,
$\square^{-1} R$ cannot begin to grow until after the onset of
matter dominance; thereafter, because of the propagator, its growth
becomes logarithmic.

The model's defining equations \cite{ourmodel} take the form,
$G_{\mu\nu} + \Delta G_{\mu\nu} = 8 \pi G T_{\mu\nu}$, with
\begin{eqnarray}
\lefteqn{ \Delta G_{\mu\nu} = \Bigl[ G_{\mu\nu} \!+\! g_{\mu\nu}
\square \!-\! D_{\mu} D_{\nu} \Bigr] \Biggl\{ f\Bigl(
\frac1{\square} \, R \Bigr) \!+\! \frac1{\square} \Bigl[ R f'\Bigl(
\frac1{\square} \, R \Bigr) \Bigr] \Biggr\} } \nonumber \\
& \hspace{3.5cm} + \Bigl[ \delta_{\mu}^{~ (\rho} \delta_{\nu}^{~
\sigma )} \!-\! \frac12 g_{\mu\nu} g^{\rho\sigma} \Bigr]
\partial_{\rho} \Bigl( \frac1{\square} \, R \Bigr) \partial_{\sigma}
\Biggl( \frac1{\square} \Bigl[ R f'\Bigl( \frac1{\square} \, R
\Bigr) \Bigr] \Biggr) \; . \label{noneqn}
\end{eqnarray}
The form of the nonlocal distortion function $f(X)$ can, unlike the
local models \cite{nof(R)}, be chosen to reproduce the $\Lambda$CDM
background cosmology exactly \cite{Tomi,Cedric,Vernov}. Indeed,
there is a simple analytic form for $f(X)$, effectively equivalent
to the numerical solution \cite{Cedric}
\begin{equation}
f(X) \approx 0.245 \Bigl[\tanh\Bigl(0.350 Y \!+\! 0.032 Y^2 \!+\!
0.003 Y^3\Bigr) \!-\! 1\Bigr] \quad , \quad Y \equiv X + 16.5 \; .
\label{analytic}
\end{equation}

Like all modified gravity theories, nonlocal cosmology can be
differentiated from general relativity with dark energy by how it
alters results in the solar system and how it affects structure
formation \cite{Yun2}. Koivisto has argued that there are no
conflicts with solar system constraints \cite{Tomi}. A recent study
of structure formation by Park and Dodelson revealed deviations from
general relativity in the 10\%-30\% range, which are interesting
because they are not currently excluded and should be observable by
the next generation of large scale structure surveys \cite{Park}.
While we await these observations, it is worth examining the
theoretical consistency of nonlocal cosmology in its own right. In
particular, how does the model behave for gravitationally bound
systems, does it possess extra degrees of freedom and is it stable?
Those are the questions we will study in sections~\ref{screen},
\ref{local} and \ref{stability}, respectively.

\section{Screening: Absence of Effects on Bound Systems}\label{screen}

In this section we discuss the issue of screening in modified
theories; $f(R)$ models suffer from the major problem that $R$
typically has the same sign for cosmology, where we want big effects
to explain the acceleration data, and for the solar system, where
significant deviations from general relativity are excluded by the
data. This has prompted the development of elaborate ``chameleon
mechanisms'' in which the extra scalar degree of freedom present in
$f(R)$ models is light in cosmological settings and heavy inside the
solar system \cite{Justin}. Nonlocal cosmology differs from $f(R)$
models in two crucial ways: there are no extra degrees of freedom to
mediate new forces; and the propagator $\square^{-1}$ acting on $R$
allows us to so define the nonlocal distortion function so that
there are no changes {\it at all} from general relativity in a
gravitationally bound system, yet without affecting the model's
predictions for cosmology. The first point will be demonstrated in
section~\ref{stability}; it is the second point which concerns us
here.

The key fact is that the scalar d'Alembertian $\square \equiv
(-g)^{-\frac12} \partial_{\mu} (\sqrt{-g} \, g^{\mu\nu}
\partial_{\nu})$ has {\it opposite signs} when acting on functions of
time than on functions of space. In the background cosmology, and
perturbations about it, the time dependence of the Ricci scalar is
stronger than its space-dependence. This means that $\square^{-1} R$
is typically negative for cosmology. Indeed, reproducing the
$\Lambda$CDM expansion history fixes the nonlocal distortion
function $f(X)$ only for negative $X$ \cite{Cedric}.

While gravitationally bound systems are not always static, their
space-dependence is generally stronger than that on time. That means
$\square^{-1} R$ is positive inside a gravitationally bound system.
Further, reproducing the $\Lambda$CDM expansion history requires
$f(0) = 0$ \cite{Cedric}. To completely annul all corrections inside
gravitationally bound systems it suffices to define $f(X) = 0$ for
all $X > 0$. Hence there is a very simple way for nonlocal models to
completely screen inside the solar system, the galaxy, or any other
gravitationally bound system, all without affecting the model's
desired behavior for cosmology.

We should comment that small values of $f(X)$ for $X > 0$ are quite
reasonable if one accepts our view that nonlocal comsology is the
gravitational vacuum polarization induced by the vast ensemble of
infrared gravitons created by primordial inflation. However, from
the purely phenomenological perspective of model building, it is
worth noting that the actual data from gravitationally bound systems
do not require anything like the severe restriction of $f(X) = 0$
for all $X > 0$. If the characteristic mass of a system is $M$ and
we observe at distance $r$, then $\square^{-1} R \sim G M/c^2r$.
This is never larger than about $10^{-6}$ for the solar system,
where observational constraints are tightest. Under the assumption
that $f(X)$ is analytic, Koivisto has shown that the best solar
system constraint only fixes the first derivative to the range $-5.8
\times 10^{-6} < f'(0) < 5.7 \times 10^{-6}$, and fixes none of the
higher derivatives \cite{Tomi}. This bound is easily met by the
simple analytic form (\ref{analytic}) which was found to reproduce
the $\Lambda$CDM expansion history \cite{Cedric}. Even giant spiral
galaxies do not have larger values of $\square^{-1} R$
--- it is about $10^{-7}$ at the Sun's orbit for our own galaxy --- and
the data are of course much less restrictive. The largest
$\square^{-1} R$ can get for an observable structure is about unity,
on the surface of a neutron star. Although high quality observations
of pulsar timing have been made, their interpretation as tests of
general relativity on the neutron star surface is complicated by
continuing uncertainty about the nuclear equation of state. Values
of $f(X)$ as large as $1/10$ at $X \approx 1$ are probably
consistent, while observation provides no constraint at all on
$f(X)$ for larger values of $X$.

\section{Local versus Nonlocal Formulations}\label{local}

Soon after our nonlocal model \cite{ourmodel} appeared, a
``localized'' version, based on two additional scalars, was proposed
\cite{Sergei,Kosh}. Briefly, it replaced the nonlocal terms in
(\ref{DL2}) by
\begin{equation}
R f\Bigl( \frac1{\square} \, R \Bigr) \sqrt{-g} \longrightarrow R
f(\Phi) \sqrt{-g} + \Psi \Bigl(\square \Phi \!-\! R\Bigr) \sqrt{-g}
\; . \label{locver}
\end{equation}
The local mechanism then, relied on two new scalars: $\Psi$ is a
Lagrange multiplier that enforces $\Phi = \square^{-1} R$ to recover
the original nonlinearity. As long as one is interested in the
inhomogeneous response of $\square^{-1} R$ to a given source of
stress-energy, and how that ultimately affects gravity, there is no
problem employing this localized version of the model. For example,
Koivisto's solar system constraint was derived using it \cite{Tomi}.
However, the localized version has severe {\it ghost} problems when
one considers the homogeneous DoF associated with the initial value
data of the two scalars, as we now show.

Consider first just the scalar, two-field sector. After a partial
integration, the off-diagonal term $\Psi \square \Phi$ is just the
difference of two diagonal free scalar Lagrangians, one of which is
therefore a ghost:
\begin{equation}
-\partial_{\mu} \Psi \partial_{\nu} \Phi g^{\mu\nu} = -\frac12
\partial_{\mu} (\Psi \!+\! \Phi) \partial_{\nu} (\Psi \!+\! \Phi)
g^{\mu\nu} + \frac12 \partial_{\mu} (\Psi \!-\! \Phi) \partial_{\nu}
(\Psi \!-\! \Phi) g^{\mu\nu} \; . \label{ghost}
\end{equation}
With our spacelike metric, the combination $(\Psi - \Phi)$ has
negative kinetic energy. (We thank G. Esposito-Farese for this
observation.)

While we have established the ghost nature of the purely scalar
sector, ours is really a three-field system; to include the
graviton, one must first perform a conformal metric rescaling to the
Einstein frame, as given in (17) of \cite{NOSZ}. As correctly stated
in \cite{NOSZ}, this implies that the necessary condition for
ghostlessness is
\begin{equation}
6 f'(\Phi) > 1 + f(\Phi) - \Psi > 0 \; . \label{impossible}
\end{equation}
No matter what we assume about the nonlocal distortion function $f$,
condition (\ref{impossible}) can never be met as long as the scalar
$\Psi$ is allowed to have arbitrary initial value data. The authors
of ref. \cite{NOSZ} actually concluded that the localized model can
be ghost-free for a period of time, but this ignores the virulence
of kinetic instabilities. There are so many excitations at large
wave number that quantum fluctuations in $\Psi$ would instantly
result in violation of (\ref{impossible}), no matter what classical
mean was imposed. Note also that even if gravity had stabilized the
ghost in (\ref{ghost}), it could not have prevented $(\Phi - \Psi)$
--- and a corresponding part of the metric field --- from developing
rapid and phenomenologically unacceptable time dependence.

We now allay the worry that this disease also infects the original
system. Clearly, (\ref{locver}) only yields (\ref{DL2}) after
discarding precisely the homogeneous scalar's solutions through
requiring that they, and their first time derivatives, vanish at the
initial time; this precisely discards their DoF! The next worry that
might arise is that perhaps these excitations could somehow appear
in the original, nonlocal form. Here general relativity itself
offers the prime example of how this danger is averted: the
``Newtonian'' third mode, after the two $g^{\rm TT}_{ij}$ gravitons,
is indeed dangerous if dynamical --- {\it but} is saved from
propagating by general relativity's constraint equations. This ---
identical --- salvation of (\ref{DL2}) will indeed be demonstrated
in the next section.

\section{Nonlocal Stability}\label{stability}

We will proceed for concreteness in a particular, synchronous,
gauge. There we will see that the nonlocal equations require the
same initial data, subject to exactly the same constraints, as
general relativity. We will also show that none of the DoF common to
the nonlinear and general relativistic terms is ever converted to
ghost stature by the nonlocal corrections, as will also be
illustrated by a simple, gauge independent, linearized treatment.
The section ends with a discussion of the extent to which our
conclusions depend upon assuming retarded boundary conditions for
$\square^{-1}$, and on the form (\ref{analytic}) for the nonlocal
distortion function.

\subsection{Synchronous gauge}\label{synchgauge}

Synchronous gauge is the coordinate frame of a system of timelike,
freely falling observers who are released from a spacelike surface
with zero initial relative velocities \cite{Lifsh}\footnote{While
this gauge has well-known problems with caustics, they are not
relevant to our treatment.}
\begin{equation}
ds^2 = -dt^2 + h_{ij}(t,\vec{x}) dx^i dx^j \; . \label{synch}
\end{equation}
The basic analysis and conclusions should apply in any gauge, as we
will see they do at linearized, kinematical level.

In synchronous gauge the covariant scalar d'Alembertian takes the
form
\begin{equation}
\square = -\partial_{t}^2 - \frac12 h^{ij} \dot{h}_{ij} \partial_{t}
+ \frac1{\sqrt{h}} \, \partial_{i} \Bigl( \sqrt{h} \, h^{ij}
\partial_j \Bigr) \; . \label{dAlem}
\end{equation}
Here and henceforth, $h^{ij}$ denotes the inverse of the spatial
metric $h_{ij}$, $h$ stands for the determinant of $h_{ij}$, and an
overdot represents differentiation with respect to time. The various
curvatures we require are
\begin{eqnarray}
R_{00} & = & -\frac12 h^{k\ell} \ddot{h}_{k\ell} + \frac14 h^{ik}
h^{j\ell} \dot{h}_{ij} \dot{h}_{k\ell} \; , \label{R00} \\
R_{ij} & = & \frac12 \ddot{h}_{ij} + \frac14 h^{k\ell} \dot{h}_{ij}
\dot{h}_{k\ell} - \frac12 h^{k\ell} \dot{h}_{ik} \dot{h}_{j\ell} +
\mbox{}^{3}R_{ij} \; , \label{Rij} \qquad \\
R & = & h^{k\ell} \ddot{h}_{k\ell} + \frac14 h^{ij} h^{k\ell}
\dot{h}_{ij} \dot{h}_{k\ell} - \frac34 h^{ik} h^{j\ell} \dot{h}_{ij}
\dot{h}_{k\ell} + \mbox{}^{3}R \qquad \label{Ricci}
\end{eqnarray}
where $\mbox{}^3R$ means, as usual, the intrinsic spatial curvature.

\subsection{Initial value data and constraints}\label{initial}

Let us first see that the nonlocal field equations (\ref{noneqn})
require the same initial value data as general relativity, namely,
the values of the 3-metric and its first time derivative at $t=0$:
$h_{ij}(0,\vec{x})$ and $\dot{h}_{ij}(0,\vec{x})$. The retarded
Green's function associated with $\square^{-1}$ is defined by the
differential equation
\begin{equation}
\sqrt{h} \, \square \, G[h](t,\vec{x};t',\vec{x}') = \delta(t \!-\!
t') \delta^3(\vec{x} \!-\! \vec{x}') \; , \label{Geqn}
\end{equation}
subject to retarded boundary conditions
\begin{equation}
G[h](t,\vec{x};t',\vec{x}') = 0 \qquad \forall \; t' > t \; .
\label{GBC}
\end{equation}
Even though we cannot solve equations (\ref{Geqn}-\ref{GBC}) for an
arbitrary 3-metric, their form clearly defines the Green's function
$G[h]$ at time $t$ using only the values of $h_{ij}$ and its first
time derivative for times less than or equal to $t$.

Because $\square^{-1} R$ is the integral $\int d^4x'
G[h](t,\vec{x};t',\vec{x}') R(x')$, we need only consider the second
time derivatives of the metric in $R$; the first time derivatives
and all spatial derivatives are shielded by the inverse differential
operator. From expression (\ref{Ricci}) we see that these second
time derivatives can be written in form
\begin{equation}
R = \partial_{t}^2 \ln(h) + \frac14 \Bigl( h^{ij} h^{k\ell} + h^{ik}
h^{j\ell} \Bigr) \dot{h}_{ij} \dot{h}_{k\ell} + \mbox{}^{(3)}R \; .
\label{Rform}
\end{equation}
Now use relation (\ref{dAlem}) to express second time derivatives in
terms of the scalar d'Alembertian
\begin{equation}
\partial_{t}^2 = -\square - \frac12 h^{ij} \dot{h}_{ij} \partial_t +
\frac1{\sqrt{h}} \, \partial_i \Bigl( \sqrt{h} h^{ij} \partial_j
\Bigr) \; . \label{dtform}
\end{equation}
We can obviously combine relation (\ref{dtform}) with (\ref{Rform})
to conclude that
\begin{eqnarray}
\lefteqn{R = -\square \ln(h) + \frac14 \Bigl( h^{ik} h^{j\ell} \!-\!
h^{ij} h^{k\ell} \Bigr) \dot{h}_{ij} \dot{h}_{k\ell} } \nonumber \\
& & \hspace{3cm} + h^{ij} \Bigl( \Gamma^k_{~ ij , k} \!+\!
\Gamma^k_{~ k i , j} \!-\! \Gamma^k_{~ k\ell} \Gamma^{\ell}_{~ ij}
\!-\! \Gamma^{k}_{~ \ell i} \Gamma^{\ell}_{~ kj} \!-\! \Gamma^k_{~
ki} \Gamma^{\ell}_{~ \ell j} \Bigr) \; . \qquad \label{Rbox}
\end{eqnarray}
Here $\Gamma^{k}_{~ij} \equiv \frac12 h^{k\ell} (h_{\ell i , j} +
h_{j\ell , i} - h_{ij , \ell})$ is the 3-space affinity, and commas
denote partial differentiation.

With relations (\ref{Geqn}-\ref{GBC}), equation (\ref{Rbox}) shows
that $\square^{-1} R$ involves only the usual initial value data,
$h_{ij}(0,\vec{x})$, and $\dot{h}_{ij}(0,\vec{x})$ of general
relativity. That these initial value data are apportioned, also as
in general relativity, between constrained fields and gravitational
radiation modes is seen by examining the nonlocal corrections
$\Delta G_{00}$ and $\Delta G_{0i}$ to the constraint equations.
Note first from (\ref{GBC}) that $\square^{-1}$ and its first time
derivative both vanish at $t=0$. Further, the nonlocal distortion
function vanishes at $t=0$. So we need only examine the two terms of
(\ref{noneqn}) in which two covariant derivatives act upon
$f(\square^{-1} R) + \square^{-1} [R f'(\square^{-1} R)]$. It is
easy to see that neither of the two combinations in the constraint
equations contains a second time derivative:
\begin{eqnarray}
g_{00} \square - D_0 D_0 & = & \frac12 h^{k\ell} \dot{h}_{k\ell}
\partial_{t} - \frac1{\sqrt{h}} \, \partial_k \Bigl( \sqrt{h} \,
h^{k \ell} \partial_{\ell} \Bigr) \; , \qquad \label{O00} \\
g_{0i} \square - D_0 D_i & = & -\partial_{t} \partial_i + \frac12
h^{k\ell} \dot{h}_{i k} \partial_{\ell} \; . \label{O0i}
\end{eqnarray}
Hence we conclude that the nonlocal corrections to the constraint
equations
\begin{equation}
t = 0 \qquad \Longrightarrow \qquad \Delta G_{00} = 0 = \Delta
G_{0i}
\end{equation}
vanish at $t=0$. This completes the verification that the nonlocal
model and general relativity share the same initial data and
constraints.

\subsection{No ghosts}\label{ghosts}

To see that there are no ghosts it suffices to examine the second
derivative terms (still in synchronous gauge) of the dynamical
equations, $G_{ij} + \Delta G_{ij} = 8\pi G T_{ij}$. The second
derivatives of $h_{ij}(t,\vec{x})$ in the Einstein tensor are, from
(\ref{Rij}-\ref{Ricci}),
\begin{equation}
G_{ij} = \frac12 \ddot{h}_{ij} - \frac12 h_{ij} h^{k\ell}
\ddot{h}_{k\ell} + O(\partial_t) \; . \label{GRkey}
\end{equation}
Of course it is only the first term, $\frac12 \ddot{h}_{ij}$, that
involves unconstrained fields; the second term represents completely
constrained ones. Because general relativity has no ghosts, we need
only check that the nonlocal corrections in (\ref{noneqn}) don't
change the sign of the $\frac12 \ddot{h}_{ij}$ term in
(\ref{GRkey}).

The work of the previous subsection shows that local second time
derivatives can only come from the parts of $\Delta G_{ij}$ which
either multiply $G_{ij}$ or have two covariant derivatives acting on
$f(\square^{-1} R) + \square^{-1} [R f'(\square^{-1} R)]$. The
latter terms,
\begin{equation}
g_{ij} \square - D_i D_j = h_{ij} \square + O(\partial_t)
\end{equation}
are simple to analyze. The local second derivative terms are
therefore,
\begin{eqnarray}
\lefteqn{G_{ij} + \Delta G_{ij} = \frac12 \ddot{h}_{ij} \times
\Biggl[ 1 + f\Bigl( \frac1{\square} \, R\Bigr) + \frac1{\square}
\Bigl[ R f'\Bigl( \frac1{\square} \, R\Bigr) \Bigr] \Biggr] } \nonumber \\
& & \hspace{-.5cm} - \frac12 h_{ij} h^{k\ell} \ddot{h}_{k\ell}
\times \Biggl[ 1 + f\Bigl( \frac1{\square} \, R\Bigr) +
\frac1{\square} \Bigl[ R f'\Bigl( \frac1{\square} \, R\Bigr) \Bigr]
- 4 f'\Bigl( \frac1{\square} \, R \Bigr) \Biggr] + O(\partial_t) \;
. \qquad \label{nonkey}
\end{eqnarray}
Only the first line of expression (\ref{nonkey}) represents the
unconstrained, dynamical part of $h_{ij}$. By comparing with the
approximate analytic form (\ref{analytic}) of the nonlocal
distortion function $f(X)$ we see that the coefficient of the
dynamical term is reduced at late times, but never by enough to make
it change sign. We therefore conclude that no dynamical graviton
mode ever becomes a ghost.

\subsection{No linearized ghosts}\label{nolin}

As a complement to our detailed treatment of DoF in the full
nonlinear theory, the present subsection is devoted to the
linearized (about flat space $g_{\mu\nu} = \eta_{\mu\nu} +
k_{\mu\nu}$) treatment of the problem.\footnote{For an analysis of
perturbations from a related nonlocal model \cite{TW3} in a general
cosmological background see \cite{TW4}.} This has several
advantages: First, it is of course simpler, yet it retains the main
point of the DoF analysis, since their content resides. Second, it
allows us to treat the desired results gauge invariantly. [Of
course, the full nonlinear treatment is needed to make sure no
higher order failure of the critical constraint equations occurs, as
notoriously happen in generic massive gravity models \cite{BD}.]

We first derive the relevant field equation; varying $R^{\rm lin}
\partial^{-2} R^{\rm lin}$ yields
\begin{equation}
\Delta G^{\rm lin}_{\mu\nu} = \Bigl( \eta_{\mu\nu} \partial^2 -
\partial_{\mu} \partial_{\nu} \Bigr) \frac1{\partial^2} \, R^{\rm lin} \equiv
\Pi_{\mu\nu} \frac1{\partial^2} \, R^{\rm lin} \qquad , \qquad
R^{\rm lin} = \Pi^{\rho\sigma} k_{\rho\sigma} \; .
\end{equation}
The transverse projector's $0\mu$ components are respectively of
zero and first order in time derivatives: $\Pi_{00} = -\nabla^2$,
$\Pi_{0i} = -\partial_0 \partial_i$, already showing these are
constraint components, as in (\ref{O00}-\ref{O0i}). For orientation,
we revert to synchronous gauge (here $k_{0\mu} = 0$), for which
$R^{\rm lin} = \partial^2 k^T - \nabla^2 \ddot{k}^{L}$. Here $k^T$
and $k^L$ are components of the usual ADM ``TT'' decomposition of a
symmetric spatial tensor $k_{ij} = k^{TT}_{ij} + \frac12
(\delta_{ij} - \partial_i \partial_j/\nabla^2) k^T + \frac12
(\partial_i k^T_j + \partial_j k^T_i) + \partial_i \partial_j k^L$;
indeed, $k^T$ is precisely the Newtonian metric of concern, while
$k^L$ is the doubly longitudinal, pure gauge, term \cite{ADM}. The
rest of the story is of course just the linearization of the results
of subsections~\ref{initial} and \ref{ghosts}.

Now we go to the linearized, but gauge invariant $R^{\rm lin}$:
\begin{equation}
R^{\rm lin} = \partial^2 k^T - C \qquad , \qquad C \equiv \nabla^2
k_{00} + \nabla^2 \ddot{k}^L - 2 k_{0i , 0i} \; .
\end{equation}
The first, gauge invariant, Newtonian term is unchanged, while the
additional (also gauge invariant) combination $C$ differs from its
synchronous gauge value only by lower time derivative terms, so the
justifications previously exhibited for that gauge simply carry over
unchanged to any frame.

\subsection{Generalized models}\label{genmod}

Here we consider how generalizations of the model would affect our
conclusions. We begin with the initial time, which can be any
instant during the epoch of radiation dominance \cite{Cedric}.
Unless the nonlocal distortion function $f(X)$ is changed from the
form (\ref{analytic}), the initial time could not be taken during
the epoch of primordial inflation because $\square^{-1} R$ behaves
there like $-4$ times the number of e-foldings
\cite{TW1}.\footnote{See \cite{TW3,TW4} for a related nonlocal model
which describes primordial inflation.} Nor would it make any
physical sense to assume such an early time because our physical
picture of the nonlocal modification is the gravitational vacuum
polarization that was built up {\it during} primordial inflation by
the continual production of infrared gravitons.

Our conclusions about the number of DoF and the initial value
constraints are independent of any assumption about the nonlocal
distortion function $f(X)$. However, our no-ghost conclusion does
require an $f(X)$ which maintains the positivity of the coefficient
$\ddot{h}_{ij}$ in equation (\ref{nonkey}). The choice (\ref{analytic})
which reproduces the $\Lambda$CDM expansion history will do this.
Because that $f(X)$ does not come near to changing the sign at the
current time, any excursion from (\ref{analytic}) which is still
consistent with constraints on the expansion history should be
acceptable. However, our no-ghost conclusion would be endangered
by an $f(X)$ which can become smaller than $-1$, or whose slope
can become too positive.

Finally we come to the question of modifying our assumption about
the use of retarded boundary conditions to define $\square^{-1}$.
Retarded boundary conditions seem very natural from our perspective
of viewing nonlocal cosmology as the gravitational vacuum
polarization that was built up from nothing during primordial
inflation. We cannot consider promoting the boundary condition to a
new DoF because this would recover the localized model, with its
fatal ghost. However, one might consider how the model looks with
some other, but definite boundary condition.

Suppose we fix the initial values of $\square^{-1} R$ and its first
time derivatives as $\Phi_0(\vec{x})$ and $\dot{\Phi}_0(\vec{x})$.
Green's Second Identity allows us to express $\square^{-1} R$ in
terms of the Green's function defined by relations
(\ref{Geqn}-\ref{GBC}),
\begin{eqnarray}
\lefteqn{\Bigl[\frac1{\square} \, R\Bigr](t,\vec{x}) = \int_{t' > 0}
\!\!\!\!\! d^4x' \sqrt{h(t',\vec{x}')} R(x')
G[h](t,\vec{x};t',\vec{x}') } \nonumber \\
& & + \int d^3x' \sqrt{h(0,\vec{x}')} \Bigl[ \Phi_0(\vec{x}')
\partial_{t'} G[h](t,\vec{x};0,\vec{x}') \!-\!
G[h](t,\vec{x};0,\vec{x}') \dot{\Phi}_0(\vec{x}') \Bigr] \; . \qquad
\label{general}
\end{eqnarray}
Of course the Green's function is not known for arbitrary
$h_{ij}(t,\vec{x})$ but the cosmological background
$h_{ij}(t,\vec{x}) = \delta_{ij} a^2(t)$ is simple enough to
analyze: spatially homogeneous contributions to $\Phi_0$ add
constants, whereas spatially homogeneous contributions to
$\dot{\Phi}_0$ behave as $\int dt/a^3(t)$.

Permitting nonzero $\Phi_0$ and $\dot{\Phi}_0$ would obviously
change our result that the initial value constraints of nonlocal
cosmology agree with those of general relativity. Nonzero values for
$\Phi_0$ and $\dot{\Phi}_0$ also change the numerical value
--- although not the functional form --- of the crucial coefficient
of the $\ddot{h}_{ij}$ term in equation (\ref{nonkey}). Although
small changes of this type pose no essential problem, neither change
is particularly desirable. So it is just as well to stick with the
original model with $\Phi_0(\vec{x}) = 0 = \dot{\Phi}_0(\vec{x})$,
which is also what the putative physical origin of the nonlocal
correction would suggest.

\section{Discussion}\label{discuss}

Our nonlocal model (\ref{DL2}-\ref{noneqn}) exactly reproduces the
$\Lambda$CDM expansion history with zero cosmological constant
\cite{Cedric}. The model has no current problem either with solar
system tests \cite{Tomi} nor with existing data on structure
formation \cite{Park}. The small deviations from general relativity 
it predicts for structure formation should be resolvable with the 
next generation of large scale structure surveys \cite{Park}. In 
anticipation, we have considered the theoretical issues of screening 
and stability.

Our first result is that screening inside any gravitationally bound
system can be made 100\% effective by simply defining the nonlocal
distortion function to vanish for positive argument, which has no
effect on the (desired) cosmological behavior. Our second result is
that the localized model \cite{Sergei} is inequivalent to ours in
that it has extra (scalar) excitations, one of which is unavoidably
a ghost. Instead, we saw that the nonlocal model has the same DoF as
general relativity; the variables of both separate into identical
sets of constrained and radiation excitation modes, subject to the
identical initial value constraints. Further, despite the (also
desired!) difference in the evolution equations, explicit and
nonperturbative examination of the highest time derivatives shows
that no graviton degree of freedom ever becomes a ghost. That
ensures the absence of kinetic energy instabilities. The more
difficult issue, ruling out instabilities due to possible negative
potential energy excitations is, if anything, more difficult than
proving the positive energy theorem in general relativity; while
these bad modes seem unlikely on physical grounds, we have not
attempted to exclude them.

{\it Note added in proof}: A very recent study \cite{Scott} now
finds that our model deviates quite significantly from observed
structure formation data. Whatever the outcome, this of course in 
no way affects our nonlocal models' inner consistency properties.

\vskip 1cm

\centerline{\bf Acknowledgements}

We thank S. Dodelson and S. Park for the persistent questions which
have prompted this study. And it is pleasure to acknowledge
stimulating conversations with J. Khoury and C. Skordis. This work
was partially supported by Department of Energy Grant
DE-FG02-16492ER40701, by National Science Foundation grants
PHY-1266107 and PHY-1205591, and by the Institute for Fundamental
Theory at the University of Florida.

\end{document}